\documentclass[reprint, amsmath, amssymb, aps, prl]{revtex4-2}

\usepackage{graphicx}
\usepackage{tikz} 
\tikzset{
  every node/.append style={
    execute at begin node={\everymath{\displaystyle}}
  }
}
\usepackage{dcolumn}
\usepackage{bm}
\usepackage{ulem}
\usepackage{xcolor}
\usepackage{slashed}
\usepackage{hyperref}
\usepackage{afterpage} 
\usepackage{microtype} 

\hypersetup{hidelinks}

\usepackage{booktabs}

\def\Tr{\text{Tr}}

\begin{document}

\title{Simple Models of Thermal Order in $2+1$ Dimensions}

\author{Michael Smolkin}
\email{michael.smolkin@mail.huji.ac.il}

\author{Lev Yung}
\email{lev.yung@mail.huji.ac.il}

\affiliation{Racah Institute of Physics, The Hebrew University, Jerusalem 9190401, Israel}

\date{\today}

\begin{abstract}
Heating need not destroy order. We show that it persists to arbitrarily high temperatures in simple quantum field theories in $2+1$ spacetime dimensions. In a $\mathbb Z_2\times\mathbb Z_2$ model of two real scalar fields, a sufficiently negative mixed quartic coupling drives one field to condense, spontaneously breaking its~$\mathbb Z_2$ symmetry. When coupled to massless Dirac fermions through a Yukawa interaction, this condensate generates a fermion mass, thereby spontaneously breaking parity and time-reversal symmetries. We extend this construction to a $\mathbb Z_2\times O(n)$ scalar model, in which the Mermin-Wagner theorem precludes spontaneous breaking of the continuous $O(n)$ symmetry at finite temperatures. Nevertheless, for $n=2$, a Berezinskii-Kosterlitz-Thouless superfluid phase persists to arbitrarily high temperatures. In this phase, spatial correlations decay algebraically, with an exponent that approaches zero as the temperature increases. For $n\geq 3$, we uncover a further unexpected feature of these models: the characteristic size of ordered domains grows asymptotically with temperature, revealing an unusual enhancement of spatial correlations upon heating. We establish our results using a controlled high-temperature expansion, with selected findings corroborated by numerical functional renormalization group analysis. The simplicity of the field content in these models provides a promising starting point for exploring experimental realizations of persistent thermal order in thin-film materials.
\end{abstract}

\maketitle

\section{Introduction}
Thermal fluctuations typically restore spontaneously broken symmetries: as temperature rises, the entropic contribution to the free energy favors disordered, symmetric phases. Although this picture underlies much of our understanding of phase transitions, it is not universal.

Counterexamples include symmetry lowering upon heating in Rochelle salt~\cite{landau1980statistical}, solidification upon heating in $^3\mathrm{He}$ through the Pomeranchuk effect~\cite{Pomeranchuk:1950}, and symmetry non-restoration in Weinberg's four-dimensional $O(N)\times O(N)$ scalar model~\cite{weinberg1974gauge}. However, the material effects occur only over finite temperature ranges, while Weinberg's theory loses perturbative control before reaching its Landau pole. These limitations raise the question of whether an ordered phase can persist to arbitrarily high temperatures.

Recent work has established persistent spontaneous symmetry breaking (PSSB) in a range of UV-complete relativistic quantum field theories~\cite{Chai:2020zgq, Chai:2020onq, Liendo:2022bmv, Chaudhuri:2020xxb, Bajc:2020gpa, Chai:2021djc, Chaudhuri:2021dsq, Chai:2021tpt, Hawashin:2024dpp, Komargodski:2024zmt, SmolkinYung:2025parity, Bajc:2026ppk, Chaudhuri:2026ges, Smolkin:2026bxa, Harder:2026jjr} with an early precursor in~\cite{Salomonson:1984px}. Lattice studies have illuminated an entropic mechanism for such order: ordering in one sector can increase the entropy of another, stabilizing the ordered phase at arbitrarily high temperatures~\cite{Huang:2025gvi, Han:2025eiw, Andriolo:2026udg}. Holographic constructions also exhibit high-temperature ordered phases~\cite{Buchel:2020thm, Buchel:2020jfs, Buchel:2021ead, Buchel:2022zxl, Buchel:2023zpe, Buchel:2025cve, Buchel:2025jup, Buchel:2026iai}, including a recent example in which the ordered phase is entropically favored~\cite{Buchel:2026iai}.

Despite this progress, connecting existing quantum field theory constructions of PSSB to experimentally accessible systems remains challenging. Many existing constructions require tuning to a multicritical point, involve many field components, or both. For example, the critical biconical $O(N)\times\mathbb Z_2$ model has three relevant directions and an estimated lower threshold $N_c\simeq15$ for PSSB in $2+1$ dimensions~\cite{Hawashin:2024dpp, Smolkin:2026wij}. A recent construction in $3+1$ dimensions likewise involves two gauge sectors, hundreds of fermion flavors, and tens of thousands of real scalar components~\cite{Harder:2026jjr}. These requirements present substantial obstacles to experimental realization.

Here, we consider a class of local, unitary, UV-complete, nonconformal quantum field theories in $2+1$ dimensions that exhibit PSSB with as few as two fields. The simple field content of these models may help guide the search for experimental realizations of persistent thermal order. In a material, the field-theory description necessarily has a finite range of validity; the experimental goal is therefore to realize order that persists up to the temperature scale at which this description breaks down. We leave the discussion of candidate thin-film materials to future~work.

We begin with a $\mathbb Z_2\times\mathbb Z_2$-symmetric theory of two real scalars. At temperatures well above the mass and quartic-coupling scales, resumming the leading thermal corrections gives a controlled perturbative expansion in $1/\ln(T/\lambda)$, where $\lambda$ denotes a characteristic quartic-coupling scale. For a sufficiently negative mixed quartic coupling consistent with stability, one field condenses, and its $\mathbb Z_2$ symmetry remains broken as $T\to\infty$. The mechanism of persistent $\mathbb Z_2$ breaking extends to $O(n)\times\mathbb Z_2$ models for every positive integer $n$. 

We next couple the condensing scalar field to massless Dirac fermions through a Yukawa interaction. The scalar condensate then spontaneously breaks parity and time reversal and generates a fermion mass. This provides a simple setting for the persistent breaking of these symmetries established in~\cite{SmolkinYung:2025parity, Smolkin:2026bxa}. Time-reversal breaking manifests itself in the anomalous Hall effect, in which an applied electric field induces a transverse current even in the absence of an external magnetic field~\cite{Nagaosa:2009ycg}. Hence, if the fermions carry electric charge, an anomalous Hall response could provide an experimental signature of this symmetry breaking.

Finally, we consider an $O(n_1)\times O(n_2)$ scalar theory with $n_1 \geq 1$ and $n_2\geq 2$. For a range of quartic couplings, the theory flows to a nonlinear sigma model in the infrared. At the matching scale, its dimensionless stiffness grows as $K \sim \ln(T/\lambda)$. The Mermin-Wagner theorem forbids spontaneous breaking of the continuous $O(n_2)$ symmetry at finite temperatures. Nevertheless, nearby spins tend to align, forming locally ordered domains. For $n_2 \geq 3$ and sufficiently negative mixed quartic coupling, the characteristic size of these domains grows asymptotically with temperature, revealing a counterintuitive tendency of spatial correlations to extend over ever larger distances upon heating. 
For $n_2 = 2$, the growing stiffness suppresses vortex proliferation and sustains a Berezinskii-Kosterlitz-Thouless (BKT) superfluid phase~\cite{Berezinskii:1970pzv, Kosterlitz:1973xp} at arbitrarily high temperatures. The $O(2)$ symmetry remains unbroken, and yet the correlations decay algebraically with distance, with an exponent $\eta\sim1/\ln(T/\lambda)$ that tends to zero upon heating. This model thus provides a simple setting for the high-temperature superfluidity in two spatial dimensions previously established in~\cite{Hawashin:2026smh, Chowdhury:2026wvs}.

\section{Persistent Breaking of $\mathbb Z_2$ Symmetry}
Consider a $d=3$ Euclidean field theory containing two real scalar fields $\phi$ and $\chi$. We begin with the following $\mathbb{Z}_2\times\mathbb{Z}_2$-symmetric action
\begin{equation}\label{eq:scalar_model}
\begin{split}
S = \int d^3 x \bigg\{
\frac12(\partial\phi)^2
+\frac12(\partial\chi)^2 +\frac{1}{2}\mu_\phi^2 \phi^2+\frac{1}{2} \mu_\chi^2 \chi^2
\\+\frac{\lambda_\phi}{8}\,\phi^4
+\frac{\lambda_{\phi\chi}}{4}\,\phi^2\,\chi^2
+\frac{\lambda_\chi}{8}\,\chi^4 \bigg\},
\end{split}
\end{equation}
where $\mu_i^2$ are the quadratic mass couplings and $\lambda_j$ are the classically relevant quartic couplings of mass dimension one. For arbitrary quadratic couplings, a bounded UV potential requires~$\lambda_\phi, \lambda_\chi>0$ and $\lambda_{\phi\chi} > -\sqrt{\lambda_\phi \lambda_\chi}$. 

The model is ultraviolet-free and super-renormalizable. At zero or low temperatures, $T\ll |\mu_i|$, it is perturbatively tractable for $|\mu_\phi|\sim|\mu_\chi|\gg|\lambda_j|$. In this regime, away from phase boundaries, the classical potential determines the pattern of symmetry breaking: the vacuum is symmetric when both $\mu_i^2$ are positive, whereas a negative squared mass destabilizes the origin and leads to an ordered phase. Depending on the coupling values, four low-temperature phases are possible: (1) SYM-SYM: $\langle\phi\rangle = \langle\chi\rangle = 0$, (2) SYM-SSB: $\langle\phi\rangle = 0$, $ \langle\chi\rangle \neq 0$, (3) SSB-SYM: $\langle\phi\rangle \neq 0$, $ \langle\chi\rangle = 0$, (4) SSB-SSB: $\langle\phi\rangle \neq 0$, $ \langle\chi\rangle \neq 0$. We now ask how the symmetry-breaking pattern changes at high temperatures at fixed couplings, and whether the resulting phase structure can be determined within a controlled approximation.

We introduce a finite temperature $T$ by compactifying the Euclidean time direction on a circle of circumference $\beta = 1/T$ and imposing periodic and antiperiodic boundary conditions on the bosonic and fermionic fields, respectively. Consequently, the temporal momentum is restricted to the discrete Matsubara frequencies, and the three-dimensional Euclidean momentum integrals are replaced by finite-temperature bosonic and fermionic sum-integrals,
\begin{equation*}
\sum_K \equiv T\sum_{n\in\mathbb{Z}}\int\frac{d^{2}\mathbf{k}}{(2\pi)^{2}}, \quad \sum_{\{K\}}
 \equiv
 T\sum_{n\in\mathbb{Z}+\tfrac{1}{2}}
 \int\frac{d^{2}\mathbf{k}}{(2\pi)^{2}},
\end{equation*}
where $K = (k_0 , \mathbf{k})$ with $ k_0 = 2\pi n T$. We now argue that, in the high-temperature limit and fixed couplings, $T\gg|\mu_i|,|\lambda_j|$, the effective action can be truncated to the following leading-order form 
\begin{align}\label{eq:effective_action}
\Gamma_\mathrm{LO}[\phi,\chi] =& \int_0^\beta d\tau\int d^{2} x [
\tfrac12(\partial\phi)^2
+\tfrac12(\partial\chi)^2 + V_\mathrm{LO}(\phi,\chi)],\nonumber\\
V_\mathrm{LO}(\phi,\chi) =\,& \frac{1}{2}M_\phi^2 \phi^2+\frac{1}{2} M_\chi^2 \chi^2\\&+\frac{\lambda_\phi}{8}\,\phi^4
+\frac{\lambda_{\phi\chi}}{4}\,\phi^2\,\chi^2
+\frac{\lambda_\chi}{8}\,\chi^4,\nonumber
\end{align}
where $M_\phi^2$ and $M_\chi^2$ are the thermal masses determined below. Higher-order corrections are suppressed either by inverse powers of
\begin{equation}
\ell\equiv\ln(T/\lambda),
\end{equation}
or, more strongly, by powers of  $|\mu_i|/T$ and $|\lambda_j|/T$. Here~$\lambda\sim|\lambda_j|$ is a characteristic scale of the quartic couplings. The thermal masses are determined by the following self-consistency equations, obtained from quartic one-loop diagrams evaluated with dressed propagators,
\begin{equation}\label{eq:self-energies}
\begin{split}
    &M_{\phi}^2 = \mu_\phi^2+ \frac{3 \lambda_\phi}{2}\sum_K G_\phi(K) + \frac{\lambda_{\phi\chi}}{2}\sum_K G_\chi(K),\\
&M_{\chi}^2 = \mu_\chi^2+\frac{\lambda_{\phi\chi}}{2}\sum_K G_\phi(K) + \frac{3\lambda_{\chi}}{2}\sum_K G_\chi(K),
    \end{split}
\end{equation}
where $G_i(K)=1/(K^2+M_i^2)$ in the symmetric phase. If~$M_i^2<0$ for either field, the symmetric configuration is unstable, and we instead use the propagators for fluctuations around the minimum $(\langle\phi\rangle$, $\langle \chi \rangle)$ of the effective potential~\footnote{We match $\Gamma^{(2)}_{{\rm LO}, ii}\big|_{\langle\phi\rangle,\langle\chi\rangle}$ at zero external momentum to the sum of 1PI two-point diagrams evaluated in the same background. The explicit $\langle\phi\rangle^2$ and $\langle\chi\rangle^2$ terms in $\Gamma^{(2)}_{{\rm LO}, ii}$ cancel against the identical terms in the tree-level diagrams. The remaining 1PI loop contributions determine $M_i^2-\mu_i^2$, giving~\eqref{eq:self-energies} at leading order.}: 
\begin{equation}\label{eq:correlators}
\begin{split}
    &\left(\begin{matrix}
        G_{\phi} & G_{\phi\chi} \\
        G_{\phi\chi} & G_{\chi}
    \end{matrix}\right) = \left.\left(\begin{matrix}
        \Gamma^{(2)}_{\phi\phi} & \Gamma^{(2)}_{\phi\chi} \\
        \Gamma^{(2)}_{\phi\chi} & \Gamma^{(2)}_{\chi\chi}
    \end{matrix}\right)^{-1}\right|_{\langle\phi\rangle, \langle\chi\rangle}^\mathrm{LO},
\end{split}
\end{equation}
with 
\begin{equation}\label{eq:1pi-two-point}
\begin{split}
        &\Gamma^{(2)}_{\mathrm{LO},\phi\phi}=K^2 + M_\phi^2 + \frac{3\lambda_\phi}{2}\phi^2+\frac{\lambda_{\phi\chi}}{2}\chi^2, \\
        &\Gamma^{(2)}_{\mathrm{LO},\chi\chi}= K^2+ M_\chi^2 + \frac{\lambda_{\phi\chi}}{2}\phi^2+\frac{3\lambda_{\chi}}{2}\chi^2,\\ &\Gamma^{(2)}_{\mathrm{LO},\phi\chi} = \Gamma^{(2)}_{\mathrm{LO},\chi\phi} = \lambda_{\phi\chi}\phi\chi.
\end{split}
\end{equation}
The self-consistency equation~\eqref{eq:self-energies} sums an infinite series of cactus diagrams with loops formed by the bare propagators around the minimum. Schematically, in the symmetric phase,
\begin{center}

\tikzset{every picture/.style={line width=0.75pt}} 

\begin{tikzpicture}[x=0.75pt,y=0.75pt,yscale=-1,xscale=1]

\draw [color={rgb, 255:red, 0; green, 0; blue, 0 }  ,draw opacity=1 ]   (168.5,134.48) .. controls (169.72,119.68) and (188.26,122.1) .. (188.23,134.48) ;
\draw [color={rgb, 255:red, 0; green, 0; blue, 0 }  ,draw opacity=1 ]   (188.23,134.48) .. controls (187.01,149.27) and (168.46,146.85) .. (168.5,134.48) ;
\draw [color={rgb, 255:red, 0; green, 0; blue, 0 }  ,draw opacity=1 ]   (191.26,88.93) -- (165.2,88.93) ;
\draw [color={rgb, 255:red, 0; green, 0; blue, 255 }  ,draw opacity=1 ]   (169.66,79.76) .. controls (170.68,67.3) and (186.3,69.34) .. (186.27,79.76) ;
\draw [color={rgb, 255:red, 0; green, 0; blue, 255 }  ,draw opacity=1 ]   (186.27,79.76) .. controls (185.24,92.21) and (169.63,90.17) .. (169.66,79.76) ;
\draw [color={rgb, 255:red, 0; green, 0; blue, 255 }  ,draw opacity=1 ]   (262.13,66.43) .. controls (263.16,53.97) and (278.79,56) .. (278.76,66.43) ;
\draw [color={rgb, 255:red, 0; green, 0; blue, 255 }  ,draw opacity=1 ]   (278.76,66.43) .. controls (277.74,78.9) and (262.1,76.87) .. (262.13,66.43) ;
\draw [color={rgb, 255:red, 0; green, 0; blue, 255 }  ,draw opacity=1 ]   (238.22,68.61) .. controls (239.25,56.14) and (254.88,58.18) .. (254.85,68.61) ;
\draw [color={rgb, 255:red, 0; green, 0; blue, 255 }  ,draw opacity=1 ]   (254.85,68.61) .. controls (253.82,81.08) and (238.19,79.04) .. (238.22,68.61) ;
\draw [color={rgb, 255:red, 0; green, 0; blue, 255 }  ,draw opacity=1 ]   (273.73,54.12) .. controls (274.75,41.65) and (290.39,43.68) .. (290.36,54.12) ;
\draw [color={rgb, 255:red, 0; green, 0; blue, 255 }  ,draw opacity=1 ]   (290.36,54.12) .. controls (289.33,66.58) and (273.7,64.55) .. (273.73,54.12) ;
\draw [color={rgb, 255:red, 0; green, 0; blue, 0 }  ,draw opacity=1 ]   (272.82,88.93) -- (246.77,88.93) ;
\draw [color={rgb, 255:red, 0; green, 0; blue, 255 }  ,draw opacity=1 ]   (251.22,79.76) .. controls (252.25,67.3) and (267.86,69.34) .. (267.83,79.76) ;
\draw [color={rgb, 255:red, 0; green, 0; blue, 255 }  ,draw opacity=1 ]   (267.83,79.76) .. controls (266.81,92.21) and (251.19,90.17) .. (251.22,79.76) ;

\draw [color={rgb, 255:red, 0; green, 0; blue, 255 }  ,draw opacity=1 ]   (345.61,66.43) .. controls (346.63,53.97) and (362.27,56) .. (362.24,66.43) ;
\draw [color={rgb, 255:red, 0; green, 0; blue, 255 }  ,draw opacity=1 ]   (362.24,66.43) .. controls (361.21,78.9) and (345.58,76.87) .. (345.61,66.43) ;
\draw [color={rgb, 255:red, 0; green, 0; blue, 255 }  ,draw opacity=1 ]   (321.7,68.61) .. controls (322.72,56.14) and (338.35,58.18) .. (338.33,68.61) ;
\draw [color={rgb, 255:red, 0; green, 0; blue, 255 }  ,draw opacity=1 ]   (338.33,68.61) .. controls (337.3,81.08) and (321.67,79.04) .. (321.7,68.61) ;
\draw [color={rgb, 255:red, 0; green, 0; blue, 255 }  ,draw opacity=1 ]   (323.87,51.22) .. controls (324.9,38.75) and (340.53,40.79) .. (340.5,51.22) ;
\draw [color={rgb, 255:red, 0; green, 0; blue, 255 }  ,draw opacity=1 ]   (340.5,51.22) .. controls (339.47,63.68) and (323.84,61.65) .. (323.87,51.22) ;
\draw [color={rgb, 255:red, 0; green, 0; blue, 255 }  ,draw opacity=1 ]   (325.32,33.83) .. controls (326.35,21.36) and (341.98,23.39) .. (341.95,33.83) ;
\draw [color={rgb, 255:red, 0; green, 0; blue, 255 }  ,draw opacity=1 ]   (341.95,33.83) .. controls (340.92,46.29) and (325.29,44.26) .. (325.32,33.83) ;
\draw [color={rgb, 255:red, 0; green, 0; blue, 255 }  ,draw opacity=1 ]   (307.04,49.06) .. controls (308.07,36.59) and (323.7,38.63) .. (323.67,49.06) ;
\draw [color={rgb, 255:red, 0; green, 0; blue, 255 }  ,draw opacity=1 ]   (323.67,49.06) .. controls (322.65,61.53) and (307.01,59.49) .. (307.04,49.06) ;

\draw [color={rgb, 255:red, 0; green, 0; blue, 0 }  ,draw opacity=1 ]   (356.29,88.93) -- (330.24,88.93) ;
\draw [color={rgb, 255:red, 0; green, 0; blue, 255 }  ,draw opacity=1 ]   (334.7,79.76) .. controls (335.72,67.3) and (351.33,69.34) .. (351.31,79.76) ;
\draw [color={rgb, 255:red, 0; green, 0; blue, 255 }  ,draw opacity=1 ]   (351.31,79.76) .. controls (350.28,92.21) and (334.67,90.17) .. (334.7,79.76) ;

\draw    (133.43,87.63) -- (122.47,87.63) ;
\draw [shift={(133.43,87.63)}, rotate = 180] [color={rgb, 255:red, 0; green, 0; blue, 0 }  ][fill={rgb, 255:red, 0; green, 0; blue, 0 }  ][line width=0.75]      (0, 0) circle [x radius= 2.01, y radius= 2.01]   ;
\draw    (145.47,87.63) -- (133.43,87.63) ;
\draw    (133.43,144.63) -- (122.47,144.63) ;
\draw [shift={(133.43,144.63)}, rotate = 180] [color={rgb, 255:red, 0; green, 0; blue, 0 }  ][fill={rgb, 255:red, 0; green, 0; blue, 0 }  ][line width=0.75]      (0, 0) circle [x radius= 2.01, y radius= 2.01]   ;
\draw    (145.47,144.63) -- (133.43,144.63) ;
\draw [color={rgb, 255:red, 0; green, 0; blue, 255 }  ,draw opacity=1 ]   (342.04,34.06) .. controls (343.07,21.59) and (358.7,23.63) .. (358.67,34.06) ;
\draw [color={rgb, 255:red, 0; green, 0; blue, 255 }  ,draw opacity=1 ]   (358.67,34.06) .. controls (357.65,46.53) and (342.01,44.49) .. (342.04,34.06) ;

\draw [color={rgb, 255:red, 0; green, 0; blue, 255 }  ,draw opacity=1 ]   (357.04,27.06) .. controls (358.07,14.59) and (373.7,16.63) .. (373.67,27.06) ;
\draw [color={rgb, 255:red, 0; green, 0; blue, 255 }  ,draw opacity=1 ]   (373.67,27.06) .. controls (372.65,39.53) and (357.01,37.49) .. (357.04,27.06) ;

\draw [color={rgb, 255:red, 0; green, 0; blue, 255 }  ,draw opacity=1 ]   (312.04,24.06) .. controls (313.07,11.59) and (328.7,13.63) .. (328.67,24.06) ;
\draw [color={rgb, 255:red, 0; green, 0; blue, 255 }  ,draw opacity=1 ]   (328.67,24.06) .. controls (327.65,36.53) and (312.01,34.49) .. (312.04,24.06) ;

\draw [color={rgb, 255:red, 0; green, 0; blue, 0 }  ,draw opacity=1 ]   (191.47,144.93) -- (165.2,144.93) ;

\draw (147.4,127.27) node [anchor=north west][inner sep=0.75pt]    {$+$};
\draw (77,70.27) node [anchor=north west][inner sep=0.75pt]  [font=\normalsize]  {$M_{i}^{2} =$};
\draw (103,130.27) node [anchor=north west][inner sep=0.75pt]  [font=\normalsize]  {$=$};
\draw (192.78,74.3) node [anchor=north west][inner sep=0.75pt]    {$+\dotsc +$};
\draw (280.11,74.66) node [anchor=north west][inner sep=0.75pt]    {$+\dotsc +$};
\draw (195.33,142) node [anchor=north west][inner sep=0.75pt]    {$,$};
\draw (369.93,74.66) node [anchor=north west][inner sep=0.75pt]    {$+\dotsc $};
\draw (147.48,74.29) node [anchor=north west][inner sep=0.75pt]    {$+$};

\end{tikzpicture}
\end{center}
where the black dot denotes the bare mass insertion $\mu_i^2$, the blue lines denote the bare propagators $1/(K^2+\mu_i^2)$, the vertices represent quartic interactions, and the black loop is evaluated using the leading-order dressed propagator. This resummation is instrumental to the perturbative argument presented below: it establishes a hierarchy among the various diagrams and ensures that higher-loop contributions are suppressed.

The thermal one-loop sum-integral evaluates to 
\begin{equation}
\begin{split}
    \sum_K \frac{1}{K^2+M^2} &=-\frac{M}{4\pi} - \frac{T}{2\pi} \ln\left(1-e^{-\frac{M}{T}}\right) \\ &= \frac{T}{2\pi}\ln \frac{T}{M} + M\,\mathcal O\left(\frac{M}{T}\right),
\end{split}
\end{equation}
where the power-law UV divergence is absent in dimensional regularization. 
Iterating~\eqref{eq:self-energies} generically yields~$|M_i^2|\sim M^2$, where we define $M^2\equiv\lambda T\ell$. 
If a thermal mass squared is negative, minimizing the effective potential~\eqref{eq:effective_action} gives a nonzero condensate whose contribution, $\lambda\langle\phi\rangle^2$ or $\lambda\langle\chi\rangle^2$, to~(\ref{eq:correlators}--\ref{eq:1pi-two-point}) is likewise of order~$M^2$. Hence, in both the symmetric and broken phases, the positive squared masses in the propagators of~\eqref{eq:self-energies} are of order $M^2$.

Using leading-order dressed propagators throughout, we show by power counting that the remaining self-energy diagrams are suppressed at high temperatures relative to the quartic one-loop diagram. Note that expanding about a nonzero condensate generates cubic vertices whose couplings $\lambda\langle\phi\rangle$ or $\lambda\langle\chi\rangle$ are of order $\lambda^{1/2}M$. Quartic and cubic vertices supply four and three ends, respectively; each internal line uses two ends, while in self-energy diagrams two ends are external. Therefore, 
\begin{equation}
    4V_4+3V_3 = 2 I + 2,
\end{equation}
where $V_4$, $V_3$, and $I$ denote the numbers of quartic vertices, cubic vertices, and internal lines, respectively. From momentum conservation in a connected diagram, the number of loops satisfies $L = I - (V_4+V_3-1)$, and hence $L=V_4+\frac{1}{2}V_3$ with $V_3$ even. 

First, retain only the zero Matsubara modes in the internal propagators. Their characteristic mass scale is $M=\sqrt{\lambda T\ell}\ll T$, and each loop carries the measure $T\int d^2k$. The only diagram that generates the IR logarithmic enhancement $\ln(T/M)$ is the quartic one-loop diagram, already accounted for in the cactus resummation defining $M_i^2$ in~\eqref{eq:self-energies}. Its insertions must therefore be excluded from all subsequent diagrams to avoid double counting. On dimensional grounds, apart from the quartic one-loop, a self-energy diagram contributes at order 
\begin{equation}
    \mbox{\small$\displaystyle
        \lambda^{V_4} (\lambda^{\frac{1}{2}}M)^{V_3} (T M^2)^L M^{-2I}
        = (\lambda T)^L M^{2-2L}
        = \lambda T\,\ell^{1-L},
    $}
\end{equation}
which is $\ll M^2$, with $\ell^{-1}$ serving as the perturbative expansion parameter. Nonzero Matsubara frequencies introduce larger gaps, $|k_0|\gtrsim T\gg M$, in the internal propagators, yielding even greater suppression. Analogous power counting shows that corrections to the terms retained in the effective action~\eqref{eq:effective_action}, including the quartic vertices, are likewise suppressed by inverse powers of $\ell$. For field amplitudes comparable to or smaller than the characteristic condensate scale $M/\sqrt{\lambda}$, the higher-power terms in the potential are $\mathcal{O}(\ell^{-1})$ relative to its quadratic and quartic terms.

\begin{figure}
    \centering
    \includegraphics[width=\linewidth]{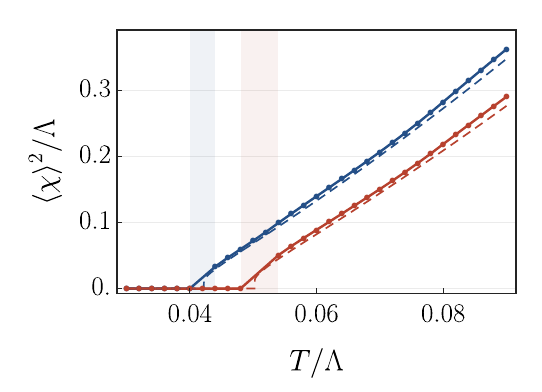}
    \par\smallskip
{\scriptsize
\setlength{\tabcolsep}{1.5pt}
\begin{tabular*}{\linewidth}{@{\extracolsep{\fill}}ccccccc@{}}
\toprule
$\mu_{\phi}/\Lambda$ &
$\mu_{\chi}/\Lambda$ &
$\lambda_\phi/\Lambda$ &
$\lambda_\chi/\Lambda$ &
$\lambda_{\phi\chi}/\Lambda$ &
$h/\sqrt{\Lambda}$ &
$d_\gamma N_f$ \\
\midrule
$2.3\times10^{-2}$ &
$10^{-4}$ &
$10^{-2}$ &
$10^{-5}$ &
$-3\times10^{-4}$ &
$4\times10^{-3}$ &
$2$ \\
\bottomrule
\end{tabular*}\par
}
    \caption{Squared thermal expectation value $\langle\chi\rangle^2$ as a function of~$T$. The solid blue and red curves show FRG results within the LPA$'8$ truncation for the scalar model~\eqref{eq:scalar_model} and its fermionic extension~\eqref{eq:fermion_action}, respectively. The shaded bands bracket the estimated critical temperatures of the inverse thermal transitions. 
    The dashed blue curve comes from the numerical solution of the leading-order thermal-mass equations~\eqref{eq:self-energies} that yields the asymptotic form~\eqref{eq:chi_tev}. The dashed red curve includes the fermionic correction~\eqref{eq:fermionic_correction_mchi}. The table lists common masses and quartic couplings, together with $h$ and $d_\gamma N_f$ for the fermionic extension. The Appendix relates the dimensionally regularized $\mu_{i}^2$ used in our analytical calculations to the FRG couplings $\mu_{i,\Lambda}^2$ used in the numerics~(\ref{eq:DR_FRG_relation}--\ref{eq:mu2_FRG}).}
    \label{fig:chi_tev_frg}
\end{figure}

The asymptotic high-temperature solution of~\eqref{eq:self-energies} admitting a controlled loop expansion takes the form
\begin{equation}\label{eq:thermal_masses_solution}
    \begin{split}
    &M_{\phi}^2 = \frac{(3 \lambda_\phi+\lambda_{\phi\chi})}{8\pi}T(\ell-\ln\ell)  + \mathcal O(\lambda T),\\
&M_{\chi}^2 = \frac{(3\lambda_\chi+\lambda_{\phi\chi})}{8\pi}T (\ell-\ln\ell) + \mathcal O(\lambda T),
    \end{split}
\end{equation}
where $\mathcal O(\lambda T)$ includes corrections of order $\mu_i^2$ as well. We take $\lambda_{\phi\chi}<0$ and choose the quartic couplings to satisfy
\begin{equation} \label{eq:lambdaconditions}
-\sqrt{\lambda_\phi \lambda_\chi}< \lambda_{\phi\chi} < -3\lambda_\chi, \qquad \lambda_\phi, \lambda_\chi >0,
\end{equation}
where the first inequality and the positivity of $\lambda_\phi, \lambda_\chi$ ensure that the effective potential is stable. The second inequality gives a negative $M_\chi^2$ at sufficiently high temperatures, inducing a nonzero thermal expectation value of the $\chi$ field~\footnote{In contrast to~\eqref{eq:thermal_masses_solution}, earlier analysis~\cite{Fujimoto:1986gu, Bimonte:1997ym} considered a solution of~\eqref{eq:self-energies} under the conditions~\eqref{eq:lambdaconditions} with both squared masses positive. As noted in~\cite{Fujimoto:1986gu}, this solution lies outside the perturbative regime with $M_\chi^2\ll \lambda T$ and thus cannot be trusted. As demonstrated in Fig.~\ref{fig:chi_tev_frg}, numerical FRG analysis supports the negative $M_\chi^2$ solution. Lattice simulations reported in Figs.~6 and~7 of~\cite{Bimonte:1997ym} use $T/\lambda_\phi < 0.12$ in our normalization and thus do not probe the high-temperature regime $T\gg |\lambda_j|$ considered in the current work.}. Minimizing the effective potential in~\eqref{eq:effective_action}, we find $\langle\phi\rangle=0$ as $T\to \infty$ and
\begin{equation}\label{eq:chi_tev}
    \langle \chi \rangle^2 = \frac{|3\lambda_\chi+\lambda_{\phi\chi}|}{4\pi\lambda_\chi}T (\ell-\ln\ell) + \mathcal O(T). 
\end{equation}
The fields $\sigma\equiv\chi-\langle\chi\rangle$ and $\phi$ describe fluctuations around either minimum. Their leading-order masses satisfy
\begin{equation}\label{eq:msigma_m_phi}
    m_\sigma^2 = - 2 M_\chi^2>0, \quad m_\phi^2 = M_\phi^2 - \tfrac{\lambda_{\phi\chi}}{ \lambda_\chi} M_\chi^2 > 0.
\end{equation}

Thus, under the conditions~\eqref{eq:lambdaconditions}, the high-temperature phase is SYM-SSB, irrespective of the $T=0$ phase. This result does not require perturbative control at zero temperature and holds without assuming a hierarchy between the scales set by $\mu_i^2$ and $\lambda_j$. When the $\chi$ sector is symmetric at low temperature, the asymptotic ordered phase requires an inverse thermal transition; when it is already ordered, the result is compatible with symmetry non-restoration. The transition temperature and critical behavior lie beyond our perturbative control, and multiple intermediate transitions are not excluded. The controlled high-temperature regime nonetheless demonstrates that the internal $\mathbb Z_2$ symmetry remains spontaneously broken at arbitrarily high temperatures.

The construction extends directly to an $O(n)\times\mathbb Z_2$ model by promoting $\phi$ to an $n$-component real scalar. For every finite integer $n\ge1$, the $\mathbb Z_2$ symmetry remains spontaneously broken at arbitrarily high temperatures, provided the quartic couplings satisfy the stability conditions and~$\lambda_{\phi\chi}<-(3/n)\lambda_\chi$.

To corroborate our analytical results, we study the model numerically using the functional renormalization group (FRG), following the setup of~\cite{Hawashin:2024dpp, SmolkinYung:2025parity}.
For an introduction to FRG, see~\cite{Wetterich:1992yh, Berges:2000ew}. Since the numerical flow starts at a finite UV cutoff $\Lambda$, we probe the high-temperature regime within the hierarchy $|\mu_{i}|, \lambda\ll T\ll\Lambda$. Our numerical results support persistent order over a range of couplings, and Fig.~\ref{fig:chi_tev_frg} shows an example of an inverse thermal transition. Increasing the scale separation gives the same qualitative results. Keeping the same quartic coupling values, but choosing small FRG masses, $\mu_{i,\Lambda}\ll|\lambda_j|$, we find an ordered phase that extends from $T=0$ to high temperatures, realizing symmetry non-restoration.
\section{Persistent Breaking of Time Reversal}
We couple the scalar model~\eqref{eq:scalar_model} to $N_f$ massless two-component Dirac fermions in the fundamental representation of $U(N_f)$ via a Yukawa interaction with $\chi$. The additional terms in the action are
\begin{equation}\label{eq:fermion_action}
    \Delta S_F = \int d^dx \left\{ \bar \psi \slashed \partial \psi + h \chi \bar\psi \psi \right\},
\end{equation}
where $h$ is the Yukawa coupling of mass dimension~$1/2$. The model remains ultraviolet free and super-renormalizable. It has a $\mathbb Z_2\times U(N_f)\times\mathcal P$ symmetry, where~$\mathbb Z_2$ acts by $\phi\to -\phi$ and $\mathcal P$ denotes the spacetime parity transformation
\begin{equation}\label{eq:parity}
\mathcal{P}:
x_1\to -x_1,\quad \chi\to -\chi, \quad \psi\to \gamma^1\psi, \quad
\bar\psi\to-\bar\psi\gamma^1.
\end{equation} 
Although coordinate directions are equivalent in Euclidean signature at $T=0$, spatial parity $\mathcal P$ and time reversal $\mathcal T$ are distinct in Lorentzian signature. In $2+1$ dimensions, the fermion bilinear $\bar\psi\psi$ is odd under both symmetries~\cite{Moshe:2003xn, Dunne1999}, so invariance of the Yukawa interaction requires $\chi$ to be odd under both as well~\footnote{For even $N_f$, the fermion flavors can be split into two groups of $N_f/2$ with opposite-sign Yukawa couplings,~$h\chi(\bar\psi^L\psi^L-\bar\psi^R\psi^R)$. Parity and time-reversal transformations may then exchange the two groups, allowing $\langle\chi\rangle\neq0$ without breaking either symmetry. This alternative model has $U(N_f/2)\times U(N_f/2)$ flavor symmetry and is not considered here. }. 

The leading-order fermionic contribution to the thermal mass of the $\chi$ field comes from the following diagram 
\begin{center}

\tikzset{every picture/.style={line width=0.75pt}} 

\begin{tikzpicture}[x=0.75pt,y=0.75pt,yscale=-1,xscale=1]

\draw    (222.1,38.57) -- (245.01,38.57) ;
\draw [shift={(245.01,38.57)}, rotate = 0] [color={rgb, 255:red, 0; green, 0; blue, 0 }  ][fill={rgb, 255:red, 0; green, 0; blue, 0 }  ][line width=0.75]      (0, 0) circle [x radius= 1.34, y radius= 1.34]   ;
\draw    (277.14,38.19) -- (300.05,38.19) ;
\draw [shift={(277.14,38.19)}, rotate = 0] [color={rgb, 255:red, 0; green, 0; blue, 0 }  ][fill={rgb, 255:red, 0; green, 0; blue, 0 }  ][line width=0.75]      (0, 0) circle [x radius= 1.34, y radius= 1.34]   ;
\draw    (245.01,38.57) .. controls (246.99,14.48) and (277.19,18.42) .. (277.13,38.57) ;
\draw [shift={(265.53,22.63)}, rotate = 180] [fill={rgb, 255:red, 0; green, 0; blue, 0 }  ][line width=0.08]  [draw opacity=0] (8.04,-3.86) -- (0,0) -- (8.04,3.86) -- cycle    ;
\draw    (277.13,38.57) .. controls (275.15,62.65) and (244.96,58.72) .. (245.01,38.57) ;
\draw [shift={(256.62,54.5)}, rotate = 6.01] [fill={rgb, 255:red, 0; green, 0; blue, 0 }  ][line width=0.08]  [draw opacity=0] (8.04,-3.86) -- (0,0) -- (8.04,3.86) -- cycle    ;

\draw (164,28) node [anchor=north west][inner sep=0.75pt]    {$\delta M{_{\chi }^{2}} \approx $};
\draw (308.2,36.39) node [anchor=north west][inner sep=0.75pt]    {$,$};

\end{tikzpicture}
\end{center}
where the arrowed line denotes the fermion propagator. At zero external momentum, this reads 
\begin{equation}\label{eq:fermionic_correction_mchi}
    \delta M_\chi^2 \approx h^2 N_f \sum_{\{K\}} \Tr \frac{1}{(i \slashed K)^2} = d_\gamma N_f\frac{\ln 2}{2\pi} h^2 T,
\end{equation}
where $d_\gamma=2$ is the dimension of the irreducible spinor representation. This contribution is subleading to the scalar result~\eqref{eq:thermal_masses_solution} as $T\to \infty$ at fixed $h$. Consequently, the nonzero $\chi$ condensate retains the leading asymptotic form~\eqref{eq:chi_tev} and induces a fermion mass $m_\psi = h \langle \chi\rangle$. This realizes persistent spontaneous breaking of parity and time reversal in the thermal state of the corresponding Lorentzian theory. The numerical FRG results shown in Fig.~\ref{fig:chi_tev_frg} corroborate this conclusion.

\section{Fate of Continuous Symmetries}
We consider an $O(n_1)\times O(n_2)$ generalization of the scalar model~\eqref{eq:scalar_model} promoting $\phi$ and $\chi$ to $n_1$- and $n_2$-component real scalars, respectively. We restrict to $n_2\geq2$ in the following. In contrast to the discrete $\mathbb Z_2$-symmetry breaking, spontaneous breaking of the continuous $O(n_2)$ symmetry at finite temperature is forbidden by the Mermin-Wagner theorem. This obstruction manifests in the $O(n_1)\times O(n_2)$ generalization of~\eqref{eq:self-energies}, where the zero Matsubara modes of the would-be Goldstone fields give an infrared-divergent one-loop correction to the squared thermal masses.

To avoid infrared divergences at this stage, we first integrate out the nonzero Matsubara modes, whose lowest gap is of order $T$, and the zero-mode fluctuations with spatial momenta above scale $M$. A controlled perturbative calculation gives the following action for the two-dimensional effective theory at scale~$M$
\begin{align}\label{eq:action_at_M}
    S_{M} =& \,\,\frac{1}{T} \int d^2x \bigg[
\frac12(\partial\phi)^2
+\frac12(\partial\chi)^2 +\frac{1}{2}M_\phi^2 \phi^2+\frac{1}{2} M_\chi^2 \chi^2 \nonumber\\&\,\,\,+\frac{\lambda_\phi}{8}(\phi^2)^2
+\frac{\lambda_{\phi\chi}}{4}\phi^2\chi^2
+\frac{\lambda_\chi}{8}(\chi^2)^2+\dots\bigg],
\end{align}
where $\phi$ and $\chi$ now refer to the dimensionally reduced fields. The dots include relative $\mathcal O(\ell^{-1})$ corrections for field amplitudes $\lesssim M/\sqrt{\lambda}$, as well as tree-level-induced momentum-dependent interactions important at~$p\sim M$. For multicomponent fields, the mass terms in~\eqref{eq:thermal_masses_solution} generalize~to
\begin{equation}\label{eq:thermal_masses_o(N_1)o(N_2)}
    \begin{split}
    &M_{\phi}^2 = \frac{((n_1+2) \lambda_\phi+n_2\lambda_{\phi\chi})}{8\pi}T(\ell-\ln\ell) + \mathcal O(\lambda T),\\
&M_{\chi}^2 = \frac{((n_2+2)\lambda_\chi+n_1\lambda_{\phi\chi})}{8\pi}T (\ell-\ln \ell) + \mathcal O(\lambda T).
    \end{split}
\end{equation}
We choose the couplings to satisfy 
\begin{equation}\label{eq:lambdaconditions_O(N1)O(N2)}
- \sqrt{\lambda_\phi \lambda_\chi}< \lambda_{\phi\chi} < -\bigl[(n_2+2)/n_1\bigr]\,\lambda_\chi,    
\end{equation}
so that the theory is stable and $M_{\chi}^2 < 0$. Now parametrize the field $\chi$ as
\begin{equation}
    \chi(x) = [v + \sigma(x)] \hat{\mathbf{s}}(x), \quad \hat{\mathbf{s}}^2 = 1, 
\end{equation}
where $\sigma$ describes radial fluctuations around $v$ and $\hat{\mathbf{s}}$ is an $n_2$-component unit vector. Here $v$ denotes the value of~$|\chi|$ at the minimum of the scale-$M$ potential. As in~\eqref{eq:chi_tev}, we obtain
\begin{equation}
    v^2 = \frac{|(2+n_2)\lambda_\chi+n_1\lambda_{\phi\chi}|}{4\pi\lambda_\chi}T (\ell-\ln\ell) + \mathcal O(T). 
\end{equation}
We integrate out the massive fluctuations of $\sigma$ and $\phi$, whose gaps are of order~$M$~\eqref{eq:msigma_m_phi}. For momenta well below both gaps, the resulting effective theory for the angular fields admits a derivative expansion,
\begin{equation} \label{eq:effective_angular_theory}
\begin{gathered}
S_{M} = \frac{K}{2}\int d^2 x \, (\partial \hat{\mathbf s})^2 + \mathcal O(\partial^4), \quad K = \frac{v^2}{T} + \mathcal O (\ell^0)\sim \ell.
\end{gathered}
\end{equation}
The infrared physics is governed by the familiar nonlinear sigma model. The stiffness $K$ controls how strongly the statistical weight favors local alignment, with $K\to0$ corresponding to uncorrelated orientations and $K\to\infty$ to strong alignment.

\textit{$n_2 \geq 3$.---}
At momenta $p\ll M$, the higher-derivative terms in~\eqref{eq:effective_angular_theory} are suppressed by powers of $p/M$. As we lower the cutoff below $M$, these terms do not generate large infrared logarithms and give only a finite $\mathcal O(\ell^0)$ correction to $K$, which we neglect. However, the two-derivative interactions arising from the constraint~$\hat{\mathbf{s}}^2=1$ induce logarithmic running of $K$~\cite{Kardar2014Lecture20_23},
\begin{equation}
    \mu\frac{d K(\mu)}{d \mu} = \frac{n_2-2}{2\pi} + \mathcal O\left(\frac{1}{K(\mu)}\right), \quad n_2 \neq 2, 
\end{equation}
where $\mu$ is the running energy scale. For $n_2\geq3$, the stiffness decreases towards the infrared, and the theory enters the strong-coupling regime. The model is known to generate a mass gap~\cite{Hasenfratz:1990ab}, resulting in a finite correlation length and no long-range order. Nevertheless, the large initial stiffness $K\equiv K(M)$ at the length scale $1/M$ implies that angular correlations extend over distances~$\xi\gg 1/M$. The correlation length is set by the scale $\mu\sim\xi^{-1}$ at which $K(\mu)$ becomes of order unity. Its high-temperature asymptotic form is given by
\begin{equation}
\begin{split}
    \xi \sim \frac{1}{ M}\,\exp\left(\frac{2\pi}{n_2-2}K + \mathcal O (\ln K)\right) \sim \frac{1}{\lambda}\left(\frac{T}{\lambda}\right)^{\Delta + o(1)},
    \end{split}
\end{equation}
where $o(1)\to 0$ as $T\to \infty$ and 
\begin{equation}
    \Delta =\frac{|(n_2+2)\lambda_{\chi} + n_1 \lambda_{\phi\chi}|}{2(n_2-2)\lambda_\chi}- \frac{1}{2}.
\end{equation}
Spins $\hat{\mathbf s}(x)$ at nearby points tend to align, forming locally ordered domains with a characteristic size set by the correlation length $\xi$. Consequently, although the Mermin-Wagner theorem forbids true long-range order at finite temperature, the size of these domains grows asymptotically with $T$ if, in addition to the conditions~\eqref{eq:lambdaconditions_O(N1)O(N2)},
\begin{equation}
    \Delta > 0 \,\,\Leftrightarrow\,\,\lambda_{\phi\chi}< - 2\bigl(n_2/n_1\bigr)\, \lambda_\chi.
\end{equation}

\textit{$n_2=2$.---} Parameterizing $\hat{\mathbf{s}} = (\cos \theta, \sin \theta)$ and restricting to smooth $\theta$ configurations, we find that the theory~\eqref{eq:effective_angular_theory} becomes Gaussian at large distances. However, the compactness of $\theta \sim \theta + 2\pi$ also permits vortex configurations with winding by $2\pi q$ around a point, where~$q\in\mathbb Z$. We represent the vortex sector using a smooth auxiliary field $\tilde \theta$, whose vertex operator $e^{iq\tilde\theta(x)}$ inserts a vortex of charge~$q$ at $x$~\cite{Tong2017}. Retaining the $q=\pm 1$ vortex operators and neglecting higher-derivative terms, we obtain
\begin{equation}\label{eq:action_atM_vortices}
    S_{1/a} \simeq \int d^2 x\Big[\frac{K}{2}(\partial \theta_\mathrm{sw})^2 + \frac{1}{8\pi^2 K} (\partial \tilde \theta)^2 - \frac{2y}{a^2} \cos \tilde\theta\Big],
\end{equation}
where $\theta_\mathrm{sw}$ is the smooth, single-valued spin-wave field. 
We lower the cutoff to $1/a$ so that $aM$ is temperature-independent and large enough for $a$ to exceed the vortex core radius. Perturbatively, this cutoff change shifts $K$ by only $\mathcal O (\ell^0)$. The vortex fugacity $y$ measures the statistical weight of a $q=\pm 1$ vortex core at resolution~$a$ in the underlying theory at scale~$M$. In the Appendix, we show that it is exponentially small in stiffness~$K$, 
\begin{equation}
    y \sim \exp[-bK + \mathcal{O}(\ln K)], \quad b > 0.
\end{equation}
Along the Gaussian fixed line~$y=0$, a vortex operator of charge~$q$ has scaling dimension~$\Delta_q=\pi Kq^2$. At sufficiently high temperatures, the stiffness~$K\sim\ln(T/\lambda)$ is large, and the fugacity~$y$ is small, so the cosine term in~\eqref{eq:action_atM_vortices} can be treated perturbatively. Since $\Delta_{\pm1}>2$ at high temperatures, the cosine is irrelevant and $y$ flows to zero in the infrared. Higher-charge vortex operators have larger scaling dimensions and are likewise irrelevant. 

The Kosterlitz-Thouless flow equations~\cite{Kardar2014Lecture20_23} show that given large $K$ and small $y$ at length scale $a$, the infrared stiffness receives an $\mathcal O (K y^2)$ correction. Since $y$ is exponentially small in $K$ as $T\to\infty$, we neglect this shift. The vortex fugacity flows to zero in the infrared, so smooth spin waves govern the long-distance physics. Their action is $S_{\rm IR}\simeq \frac{K}{2}\int d^2x (\partial\theta)^2$, where $\theta$ now denotes the spin-wave field. Using $1/a$ as the UV cutoff scale, the correlation function of $\theta$ up to a constant is given by
\begin{equation}
    \langle \theta(x)\theta(0) \rangle \simeq - \frac{1}{2\pi K} \ln \frac{|x|}{a}, \quad |x|/a\gg1.
\end{equation}
In complex notation, the infrared form of the original field is $\chi_{\text{\tiny IR}}(x) =  [v  +\mathcal O({\sqrt{T/\ell}})]\,e^{i\theta(x)}$. Using the Gaussian identity~$\langle e^{i\theta(x)} e^{-i\theta(0)} \rangle = e^{-\langle (\theta(x)-\theta(0))^2 \rangle/2}$, we obtain its long-distance two-point function at asymptotic temperatures
\begin{equation} \label{eq:BKT_correlation}
    \langle\chi(x) \chi^* (0) \rangle \simeq v^2\left(\frac{a}{|x|}\right)^{\eta},\quad \eta = \frac{1}{2\pi K},
\end{equation}
where $|x|/a\gg1$. The algebraic decay of the $\chi$ correlator identifies the Berezinskii-Kosterlitz-Thouless phase at high temperatures. This phase exhibits quasi-long-range order, with the $O(2)$ symmetry unbroken as required by the Mermin-Wagner theorem.
In the conventional two-dimensional XY model, heating unbinds vortex-antivortex pairs and destroys this order. Here the stiffness grows with temperature and suppresses vortex proliferation, so the BKT phase persists to arbitrarily high temperatures. The squared local field amplitude and the decay exponent scale as
\begin{equation}
    v^2 \sim T \ln(T/\lambda), \quad \eta \sim 1/\ln(T/\lambda), \quad T\to \infty.
\end{equation}
Thus, as the system is heated, the local amplitude grows and correlations decay ever more slowly with distance. 

Taking the exact $T=\infty$ limit at fixed $|x|$ and using $a\sim[\lambda T\ln(T/\lambda)]^{-1/2}$ in~~\eqref{eq:BKT_correlation}, we obtain
\begin{equation}
    \lim_{|x|\to \infty}\lim_{T\to\infty}\langle\chi(x) \chi^* (0) \rangle/v^2 = \exp\left\{- \frac{\lambda_\chi}{|4\lambda_\chi + n_1 \lambda_{\phi\chi}|}\right\}.
\end{equation}
Reversing the two limits gives zero. Thus, the order is quasi-long-range at every sufficiently high finite temperature, while true long-range order emerges in the strict high-temperature limit.

\section*{Acknowledgments}

This work was supported in part by BSF Grant No.~2022113, ISF Grant No.~2526/25, NSF-BSF Grant No.~2022726, and Israel’s Council for Higher Education.

\let\originalbibsection\bibsection
\renewcommand{\bibsection}{%
  \originalbibsection
  \vspace{-14pt} 
  {\centering\small\itshape
   Science is elegant.\par}
  \nobreak\vspace{4pt}
}
\nocite{Fujimoto:1986gu,Bimonte:1997ym}
\bibliography{bibliography}
\clearpage

\appendix   
\setcounter{equation}{0}
\renewcommand{\theequation}{A\arabic{equation}}

\section{Vortex Fugacity} \label{app:vortex_fugacity}
\begin{figure}
    \centering
\includegraphics[width=0.9\linewidth]{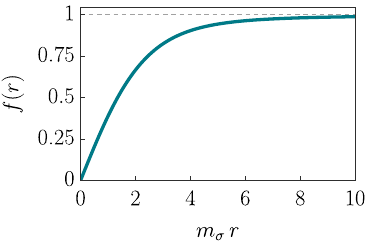}
    \caption{Radial profile $f(r)$ of a unit-winding vortex. At $r=10\,m_\sigma^{-1}$, $f(r)\simeq 0.989$, close to its asymptotic value of unity.}
    \label{fig:f_r}
\end{figure}

We use the theory at scale $M$~\eqref{eq:action_at_M} to estimate the vortex fugacity $y(a)$ at the matching scale $a$, which is a short-distance cutoff of the vortex description in~\eqref{eq:action_atM_vortices}. Consider a $q=1$ vortex configuration parametrized by
\begin{equation}
    \chi = v f(r) (\cos \varphi, \sin \varphi), \quad f(0) = 0,\quad f(\infty) = 1,
\end{equation}
where $r$ and $\varphi$ are the radial and angular space coordinates, respectively. The boundary conditions ensure regularity at the core and an energetically preferable $|\chi|$ value at large distances. For a fixed radial profile $f(r)$, uniform winding minimizes the angular-gradient energy. Setting $\phi = 0$ and substituting the $\chi$ parametrization into~\eqref{eq:action_at_M} we get up to an irrelevant constant 
\begin{equation}
\begin{split}
    S_\mathrm{vortex} &\stackrel{\mathrm{LO}}{=} \frac{1}{T} \int d^2x \bigg[\frac12(\partial\chi)^2
+\frac{\lambda_\chi}{8}(\chi^2-v^2)^2\bigg] \\
&= \frac{2\pi}{T} \int^R_0 r dr \bigg[ \frac{v^2}{2} f'^2 + \frac{v^2}{2r^2}f^2 + \frac{\lambda_\chi v^4}{8} (f^2-1)^2
\bigg], 
\end{split}
\end{equation}
where $R$ is an IR regulator. The $r\to0$ integration region remains accessible, since approaching the center of a smooth vortex does not mean probing progressively shorter wavelengths $\ll M^{-1}$. Minimizing the action with respect to $f$, we get the following equation
\begin{equation}
    f'' +f'/r -f/r^2 - \frac{m_\sigma^2}{2}(f^2-1)f=0,
\end{equation}
where $m_\sigma^2 = \lambda_\chi v^2$. We solve this equation numerically with the stated boundary conditions. The resulting vortex profile, shown in Fig.~\ref{fig:f_r}, varies on the scale~$r_{\rm core}\sim m_\sigma^{-1}$~\footnote{To better resolve the vortex core, one can raise the cutoff in~\eqref{eq:action_at_M} to a large, temperature-independent multiple of $M$. This shifts the squared thermal masses only at subleading order. The momenta $p\sim m_\sigma\sim M$ that govern the core profile then lie well below the cutoff, so the omitted momentum-dependent interactions in~\eqref{eq:action_at_M} give only small corrections to the vortex action.}.
Substituting the numerical solution into the action, we obtain
\begin{equation}
    S_{\mathrm{vortex}} = \pi K \left[\ln(R m_\sigma) + 0.034 + O((Rm_\sigma)^{-2})\right] .
\end{equation}
Subtracting the logarithmic exterior contribution $\pi K \ln(R/a)$ and taking $R\to\infty$, we get the action of the vortex core at the matching scale~$a$,
\begin{equation}\label{eq:S_core}
    S_\mathrm{core}(a) = \pi K \left[0.034 +\ln(a m_\sigma)\right].
\end{equation}
The logarithmic exterior action in the point-vortex description~\eqref{eq:action_atM_vortices} assumes $f(r)\simeq 1$ for $r\geq a$. We therefore take $a m_\sigma$ fixed as $T\to\infty$ and large enough that this approximation holds. The core action is already positive at $a=m_\sigma^{-1}$. 
Large $K$ suppresses fluctuations around the vortex saddle solution. The statistical weight $y$ of a vortex core is hence given by 
\begin{equation}
    \begin{split}
    y(a) &\simeq e^{-S_\mathrm{core}(a) + \mathcal O (\ln K)}\\ &= e^{-bK + \mathcal O (\ln K)} =\left(\frac{\lambda}{T}\right)^{c+o(1)}, \quad T \to \infty,
    \end{split}
\end{equation}
where $b, c>0$ and $o(1)\to 0$ as $T\to \infty$. Thus, the fugacity at scale $a$ tends to zero as the temperature increases. 

As we lower the cutoff from $M$ to $1/a$, neutral defects, such as vortex–antivortex pairs, can renormalize $K$ through their polarization. We expect the saddle-point profiles for well-separated cores to vary on the scale~$m_\sigma^{-1}$ and the corresponding actions to take the form~$\tilde c K$, with~$\tilde c>0$ independent of temperature. Such defects therefore have exponentially small statistical weight,~$e^{-\tilde c K}$. At resolution $M^{-1}$, the number of possible positions for defects in a region of area $a^2$ is of order $(Ma)^2$. Since this factor is fixed while $K$ grows with temperature, these defects' contribution to the running of $K$ vanishes as~$T\to\infty$.

\section{Matching DR and FRG}
In the numerical FRG calculation, the RG flow starts at a finite cutoff $\Lambda$, and the one-loop mass corrections contain terms linear in $\Lambda$. Our analytic calculations use dimensional regularization (DR), which discards such power-law divergences. The same physical theory therefore corresponds to different mass parameters in the two schemes. We consider the $O(n_1)\times O(n_2)$ scalar theory at zero temperature. For $n_2 = 1$ we allow the Yukawa interaction~\eqref{eq:fermion_action} with $N_f$ massless fermions. We relate the DR mass parameters $\mu_{i,\mathrm{DR}}^2$ to the FRG couplings $\mu_{i,\Lambda}^2$ by requiring that the quadratic terms in the effective action agree in both schemes.

The FRG infrared regulator replaces the momentum-space scalar kinetic term $\tfrac12 q^2$ by $\tfrac12 (q^2 + R_k(q))$ and fermionic $i\slashed q$ by $i\slashed q \sqrt{1+R_k(q)/q^2}$. Our numerical analysis uses the linear regulator $R_k(q) = (k^2-q^2)\theta(k^2 - q^2)$~\cite{Litim:2001up}. The difference between the infrared quadratic coupling $\mu_{i,0}^2$ in the effective action and the bare coupling $\mu_{i,\Lambda}^2$ at scale $\Lambda$ is given by the difference of quantum mass corrections evaluated with $R_{0}$ and~$R_\Lambda$, respectively. In the $\Lambda \gg |\mu_i|, |\lambda_j|, h^2$ limit, one-loop diagrams with massless propagators suffice to determine the leading $\Lambda$ dependence of $\mu_{i,0}^2 - \mu_{i,\Lambda}^2$. The resulting one-loop integral difference reads
\begin{equation}
\int_{q\leq\Lambda}\frac{d^3q}{(2\pi)^3}
\left(\frac{1}{q^2}-\frac{1}{\Lambda^2}\right)
=\frac{\Lambda}{3\pi^2}.
\end{equation}
In DR, the massless one-loop vanishes. Requiring the two schemes to reproduce the same quadratic terms in the effective action fixes the difference between the input mass couplings, 
\begin{equation}\label{eq:DR_FRG_relation}
\begin{split}
\mu_{\phi,\mathrm{DR}}^2
=\,\,& \mu_{\phi,\Lambda}^2
+\frac{(n_1+2)\lambda_\phi+n_2\lambda_{\phi\chi}}{6\pi^2}\Lambda + \dots\,,\\
\mu_{\chi,\mathrm{DR}}^2
=\,\,& \mu_{\chi,\Lambda}^2
+\frac{(2+n_2)\lambda_\chi+n_1\lambda_{\phi\chi}}{6\pi^2}\Lambda \\ -\,\,&\frac{d_\gamma N_f h^2\delta_{n_2, 1}}{3\pi^2}\Lambda + \dots\,,
\end{split}
\end{equation}
where the Yukawa term in $\mu_{\chi,\mathrm{DR}}^2$ comes from the linearly divergent $T=0$ part of the fermionic one-loop in~\eqref{eq:fermionic_correction_mchi}. The omitted terms include inverse-power cutoff corrections and higher-order contributions in the quartic and Yukawa couplings, including two-loop terms of order $\sim (\lambda^2+\lambda h^2+h^4)\ln(\Lambda/\bar\mu)$, where $\bar\mu$ is the renormalization scale in DR. At fixed couplings and scale $\bar{\mu}$, the logarithmic and finite $\mathcal O(\lambda^2, \lambda h^2, h^4)$ terms are subleading to the linear $\Lambda$ term as $\Lambda\to\infty$.

The quartic and Yukawa couplings agree between the two schemes up to inverse-power cutoff terms, as the loop corrections to the corresponding interaction vertices are UV convergent.

For the DR masses $\mu_{\phi,\mathrm{DR}} = 2.3\times10^{-2}\Lambda$ and $\mu_{\chi,\mathrm{DR}} = 10^{-4}\Lambda$ used in Fig.~\ref{fig:chi_tev_frg}, the corresponding FRG inputs are obtained via~\eqref{eq:DR_FRG_relation}, \begin{equation} \label{eq:mu2_FRG}
\begin{aligned} 
\mu_{\phi,\Lambda}^2 &= 2.74601\times10^{-5}\Lambda^2, \\ \mu_{\chi,\Lambda}^2 &= \Lambda^2 \begin{cases} 
4.56945\times10^{-6} & \text{(scalar model)}, \\ 5.65021\times10^{-6} & \text{(fermionic extension)}. 
\end{cases} \end{aligned} 
\end{equation}

\end{document}